\documentclass[aps,twocolumn]{revtex4-1}
\usepackage{amssymb, graphicx}
\usepackage{amsmath} % to have \eqref work
\usepackage{wrapfig}

\newcommand{\ket}[1]{\ensuremath{|#1\rangle}}
\newcommand{\bra}[1]{\ensuremath{\langle #1|}}
\newcommand{\G}{\mathcal{G}}

\begin{document}

\title{Multi-fractal Geometry of Finite Networks of Spins}

\author{Paul Bogdan}
\affiliation{Ming Hsieh Department of Electrical Engineering,
                 University of Southern California, Los Angeles, CA 90089},
\author{Edmond Jonckheere}
\affiliation{Ming Hsieh Department of Electrical Engineering,
                 University of Southern California, Los Angeles, CA 90089},
\author{Sophie Schirmer}
\affiliation{College of Science (Physics), Swansea University,
                 Singleton Park, Swansea SA2 8PP, UK}

% probably not needed here
%\significancetext{Quantum spin networks exploit the spin degrees of freedom to encode, process, store and transfer information. A fundamental roadblock in making this emerging technology a reality across devises of many scales is the need for a broad understanding of their information transfer capacity  across many structures (chains and rings), of varying sizes, and across many spin biasing and coupling strength manipulations for selective transfers. This is accomplished via a novel fractal geometry framework for analyzing the information transfer scaling properties of finite spin networks. Using a novel box counting argument, it demonstrates that the informational topological embedding of finite spin networks display multi-fractal properties. A thermodynamical formulation of the  multi-fractal characterization  allows the study of informational phase transition phenomena in quantum spin networks,  which could prove fundamental for the design of future networking devices.}

\begin{abstract}
{Quantum spin networks overcome the challenges of traditional charge-based electronics by encoding the information into spin degrees of freedom. Although beneficial for transmitting information with minimal losses when compared to their charge-based counterparts, the mathematical formalization of the information propagation in a spin(tronic) network is challenging due to its complicated scaling properties.  In this paper, we propose a geometric approach---specific to finite networks---for unraveling the information-theoretic phenomena of spin chains and rings by abstracting them as weighted graphs, where the vertices correspond to the spin excitation states and the edges represent the information theoretic distance between pair of nodes. The weighted graph representation of the quantum spin network dynamics exhibits a complex self-similar structure (where subgraphs repeat to some extent over various space scales). To quantify this complex behavior, we develop a new box counting inspired algorithm which assesses the mono-fractal versus multi-fractal properties of quantum spin networks. Besides specific to finite networks, multi-fractality is further compounded by ``engineering" or ``biaising" the network for selective transfer, as selectivity makes the network more heterogeneous. To demonstrate criticality in finite size systems, we define a thermodynamics inspired framework for describing information propagation and show evidence that some spin chains and rings exhibit an informational phase transition phenomenon, akin to the metal-to-insulator phase transition in Anderson localization in finite media.} 
\end{abstract}

\maketitle

Many fundamental particles such as electrons, protons and certain atomic nuclei exhibit a fundamental quantum property called spin. Spin degrees of freedom have played an important role since the discovery of nuclear magnetic resonance~\cite{Lauterbur1957} and electron spin resonance~\cite{Bagguley1947}, which have become essential tools for characterizing chemical structure, material properties and bio-medical imaging~\cite{NMR-Ernst1990, ESR-Lund2011,MRI-McRobbie2006}.  More recently, spin degrees of freedom have been in the spotlight again as potential carriers of quantum information, and the foundation of quantum spintronics~\cite{awschalom_quantum_2013,spintronics-2013}.

Conventional electronics, while powerful, also has drawbacks.  Electrical resistance encountered by moving electrons generates heat, wasting energy and limiting integration densities and data processing speeds in conventional semiconductor devices~\cite{spintronics-2001}. Spintronics in its most basic form is about exploiting spin degrees of freedom, usually of electrons, to encode, process, store and transfer information.   Encoding information in spin degrees of freedom such as excitations of a spin network opens up many possibilities including quantum information processing~\cite{QIP2015,spin-logic-2011}, where spintronics devices offer benefits such as generally long coherence lifetimes of spins at low temperatures.  Although there are many technological challenges that remain to be solved, considerable efforts are currently under way to realize various types of spintronic devices~\cite{awschalom_quantum_2013,spintronics-2013}.  For example, spin polarized currents injected at the source of the ``spin transistor" can have their polarization in the bulk of the device controlled by the gate field. 

The spintronic property that propagation happens without matter or charge transport makes spintronic networks potentially attractive for more efficient on-chip interconnectivity via ``spin channels'' even for classical information processing. In this context one of the most important questions is the capacity of a spin(tronic) network for information transport or teleportation of quantum states between nodes in the network.  One measure introduced to capture the intrinsic ability of a quantum network to transport information between nodes through the propagation of excitations is Information Transfer Fidelity (ITF)~\cite{Jonckheere2014,Jonckheere,IEEE-CCA2011,QIP2015,IEEE-CDC2015-preprint}.  Broadly, it is an easily computable upper bound on the maximum achievable probability with which an excitation can be successfully propagated from one node to another in the network.  

The ITF induces a kind of metric for the spin network that endows it with an information topology and quantifies the information-theoretic correlations between nodes in the network.  This information topology of the spin network generally differs substantially from the physical geometry of the network. We show that the node-to-node interactions in such ``symmetric" spin networks as long chains and rings exhibit self-similarity characterized by a narrow fractal spectrum, indicating mono-fractal behavior.  Shorter chains tend to have broader fractal spectra.  However, the multi-fractal property takes its full significance when the chain or the ring is manipulated---in a way that affects symmetry---to favor specific transmissions. 

Couplings in spin chains are ``engineered" to achieve perfect state transfer between end points~\cite{PhysRevA.71.032312,Christandl2004,doi:10.1142/S0219749910006514}.  A particular type of engineered chain that has been studied before is a chain with nearest-neighbor couplings satisfying~\cite{Christandl2004}
\begin{equation}
    J_{k,k+1} = \tfrac{1}{2} \sqrt{ k (N-k)}, \quad k=1,\ldots,N-1.
\end{equation}
A chain with couplings like this can be shown to exhibit perfect state transfer between antipodal points in the chain at certain
times~\cite{Christandl2004}. Rings on the other hand are biased to ``quench'' the ring to a chain hence favoring transfer near the biased spin~\cite{QIP2015,IEEE-CDC2015-preprint}.  In particular, we contrast original  and engineered spin chains and original and biased spin rings.  For both spin chains and rings, we find that the degree of multi-fractality varies with network sizes. Engineered spin chains display a more pronounced multi-fractal behavior than the original counterparts. Along the same lines, we observe that the degree of multi-fractality for spin rings is influenced by the considered bias magnitude.

The paper is organized as follows. We begin with the information geometry embedding of quantum spin networks.  We provide the basics of the mathematical description of the Information Transfer Fidelity between quantum spin excitation states.  Building on this background, next, we summarize a few challenges that an information-theoretic approach to quantum spin networks faces and highlight the possibility of analyzing the information propagation and the long-range interactions via a novel information geometry graph embedding.  We present the multi-fractal characteristics of this information geometry embedding  and a greedy algorithm for investigating the multi-fractality of quantum spin networks,  along with a novel box counting measure different from the popular one~\cite{anderson}.  In the next two sections, we detail our multi-fractal analysis of quantum chains and quantum rings, respectively, under various network sizes and various manipulations to favor some selective transfers.   We conclude the paper by outlining our main findings, discussing the deeper significance of our results, and indicating several future research directions.

\section{Information Geometry Embedding of Spin Networks}
\label{sec:ITF}

A major difficulty in constructing a rigorous information-theoretic understanding of spin networks is embodied in the foundational features of quantum mechanics.  If we encode information in the states of a quantum system, such as a network of coupled spin-$\tfrac{1}{2}$ particles, then the transfer of information between quantum states is probabilistic and governed by the laws of quantum mechanics, specifically the Schr\"odinger equation,
\begin{equation}
  \label{eq:SE}
  \imath\hbar \tfrac{d}{dt}\ket{\psi(t)} = H \ket{\psi(t)}, \quad \ket{\Psi(0)}=\ket{i}, 
\end{equation}
or a suitable open-system generalization~\cite{Breuer2002}.   In the preceding, $\ket{i}$ denotes the state where the only single excitation in the network is on spin $i$. More specifically, the evolution is characterized by a Hamiltonian $H$, a Hermitian operator with eigendecomposition
\begin{equation}
  \label{eq:H}
  H = \sum_{k=1}^N \lambda_k \Pi_k,
\end{equation}
where the $\lambda_k$'s are the (distinct) real eigenvalues and the $\Pi_k$'s are the corresponding projectors onto the corresponding eigenspaces.  Although  capturing the $t^*$ that yields maximum fidelity $\sup_{t\geq 0} |\langle j | \exp(-\imath H t/\hbar)|i\rangle|$ in the temporal evolution of a given input state $\ket{i}$ is complicated, we can derive an easily computable upper bounds on the probability of transfer of information to another state $\ket{j}$ in a network governed by the Hamiltonian $H$,
\begin{equation}
  \label{eq:p}
 \sup_{t\geq 0} |\langle j | \exp(-\imath H t/\hbar) |i\rangle|^2 \leq p_{\max}(i,j)  := \sum_{k=1}^N |\bra{j}\Pi_k \ket{i}|^2. 
\end{equation}
Besides computational easiness, the right-hand side upper bound is independent on 
the details of the capture of the wave function $\Psi(t)$ around $\ket{j}$ by, say, Anderson localization. 
Conditions for attainability of the bound in homogeneous chains and rings are derived in~\cite{Jonckheere2014}  and~\cite{QIP2015}, 
respectively. 
Taking the logarithm of the transition probability, we can define with a slight abuse an
Information Transfer Fidelity (ITF) ``distance"
\begin{equation}
   \label{eq:d}
  d(i,j) = - \log p_{\max}(i,j).
\end{equation}
Note that $p_{\max}(i,i)=1$ for any state $\ket{i}$ as $\sum_{k=1}^N\Pi_k$ is a resolution of the identity, and also that $p_{\max}(i,j)=p_{\max}(j,i)$.  Hence $d(i,i)=0$ and $d(i,j)=d(j,i)\geq 0$, so that $d(\cdot,\cdot)$ is a symmetric prametric~\cite[p. 23]{Aldrovandi1999}. Although this prametric need not be separating~\cite[p. 23]{Aldrovandi1999}, that is, $d(i,j)=0$ need not imply $i=j$ as it happens for anti-podal points of $N$ even spin rings, the latter is easily fixed by identifying those $d(i,j)=0$ points. After this identification, the resulting semi-metric  in general does not satisfy the triangle inequality, but for certain types of homogeneous networks it has been shown to induce a proper metric~\cite{Jonckheere,QIP2015}.
For the purposes of our analysis here a semi-metric (which by definition satisfies all axioms except the triangle inequality) is sufficient.

In what follows, we set forth a geometrical approach to the spin network in the sense that the information interactions within the spin network are represented as a weighted graph $\G=(\mathcal{V},\mathcal{E})$, where the vertices represent spin excitation states and the edges denote their information-theoretic semi-metric,  or simply distance~\footnote{To make the exposition more crisp, we will for here on refer to $d(\cdot,\cdot)$  as a distance, with the warning that this is an abuse of language as in general  $d(\cdot,\cdot)$ is only a semi-metric.},  as determined by Eq.~(\ref{eq:d}).  We stress here that the resulting graph representation differs from the common ``Hamiltonian'' graph representation, where the vertices are also quantum states but the edges are given by the strength of the Hamiltonian coupling between the quantum states rather than their information-theoretic distance. There is no simple relationship between these two distinct graph representations of the network.  Our graph approach contributes not only to a spatial representation of node-to-node interactions, but also captures non-local correlations by encapsulating their information-theoretic characteristics within edge weights.

Although our approach and the mathematical techniques employed are general and can be applied to any spin network, in this work we focus on simple networks such as linear arrangements (chains) or circular arrangements (rings) of spins with nearest-neighbor interactions, for which the $J$-coupling matrix is either tridiagonal (chain) or circulant (ring).  For a  network with uniform coupling all non-zero entries in the $J$ matrix are the same, and we can choose units such that they are $1$.  More generally, we can always choose units such that the maximum coupling strength is $1$.  If the dynamics of the network are restricted to the single excitation subspace then the Hamiltonian on this subspace is determined by the $J$-coupling matrix; for Heisenberg coupling there are additional non-zero elements on the diagonal, while the diagonal elements are zero for $XX$-type coupling~\cite{Bose2007}.

\section{Fractal Analysis}

\begin{figure*}[t]
\centerline{\includegraphics[width=7.2in,keepaspectratio]{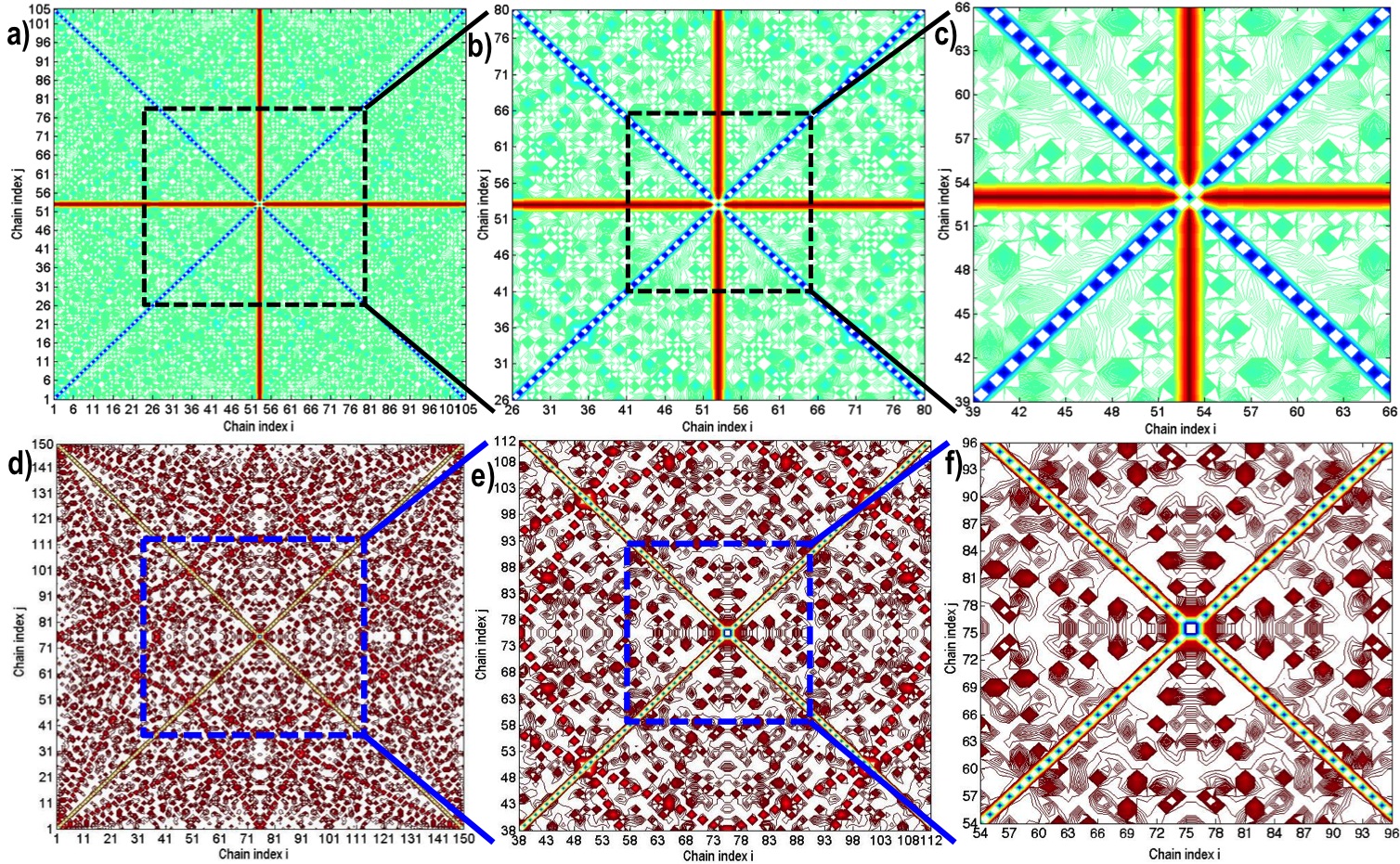}}
\caption{(a) Realization of the ITF for a spin chain of size $N=105$.
  (b) Zoom-in magnification on the ITF metric graph by a factor of $~2$.
  (c) Zoom-in magnification on the ITF metric graph by a factor of $~4$.
  (d) Realization of the ITF for a spin chain of size $N = 150$.
  (e) Zoom-in magnification of the ITF graph realization by a factor of $~2$.
  (f) Zoom-in magnification of the ITF graph realization by a factor of $~4$.}
\label{f:SpinNetworksSelfSimilarityIslands}
\end{figure*}

One important characteristic exhibited by the information distance graph representation of spin networks is the \emph{self-similarity} of the node-to-node interactions. In mathematical context, the self-similarity implies that an object (process) is exactly or approximately similar to a subcomponent under the magnification operation (the whole resemblance in shape to subcomponents).  Fig.~\ref{f:SpinNetworksSelfSimilarityIslands} shows a visual representation of the information distance for spin chains of size $N=105$ and $ N=150 $, respectively.  Although different network sizes exhibit different spatial interaction patterns, the metric graph representation is not entirely irregular under the magnification operation; rather it exhibits repetition and symmetry---there are information valleys surrounded by hill tops that seem to repeat almost identically across space, yet are not exactly identical.  From a mathematical perspective, we know that this irregularity cannot be understood by simply defining the embedding dimension as the number of variables and coordinates as considered in~\cite{IEEE-CCA2011}, but rather calls for quantifying the dimension using multi-fractal geometry \cite{mandelbrot}.

\begin{figure*}[t]
\centerline{\includegraphics[width=7.2in,keepaspectratio]{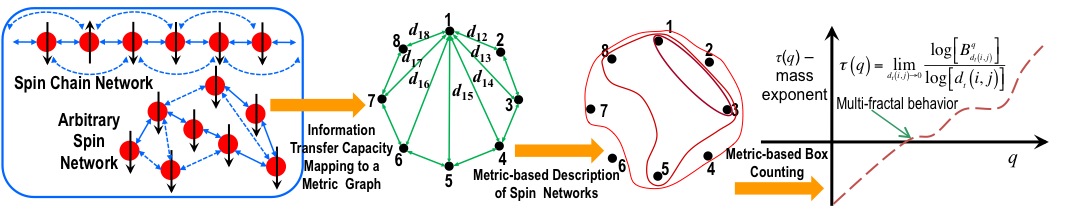}}
\caption{An arbitrary spin network with a set of heterogeneous
  coupling parameters can be represented using information-theoretic
  measures as a weighted graph. Depending on the time dependent
  probability of transfer of excitation from spin $|i\rangle$ to
  $|j\rangle$ and the information theoretic measure defined on these
  node-to-node interactions, some nodes may reside in a smaller
  geodesic island even though spatially they reside at a much larger
  physical distance. Relying on the information theoretic measure, a
  box counting inspired strategy can help to investigate the scaling
  behavior of the mass exponent and derive the multi fractal spectrum
  associated with node-to-node interactions in a spin network.}
\label{f:MetricBasedDescription}
\end{figure*}

To investigate the multi-fractal characteristics of spin networks, we adopt the following strategy (see Fig.~\ref{f:MetricBasedDescription} for a graphical representation of the strategy): First, we map the spin network dynamics onto a metric space graph, where the weights of the graph edges represent the information distance $d(i,j)$ between excitation states $|i\rangle$ and $|j\rangle$, respectively. As shown in Fig.~\ref{f:MetricBasedDescription}, this information metric-based graph representation of the spin network can also be seen as a map of contour lines (isolines), where two nodes connected by an information distance (weight) less than or equal to $d(i,j)$ belong to an island (bounded by a closed contour line) encompassing all nodes within the same $d(i,j)$ distance.  However, this is just an alternative mapping representation and the formalism is entirely developed to work with any weighted graph.  Second, we construct a graph-based box-covering renormalization inspired method \cite{mandelbrot, PhysRevA.45.6989}, which aims at covering the metric graph with a minimum number of boxes $B_{k}(\epsilon)$ of radius~\footnote{Throughput the explanation of the multi-fractal analysis, we use the concept of radius to denote all nodes which can be reached within one hop search because their distances (weights) to the center node under investigation is less than the specific radius.} $\epsilon=d(i,\omega)$ for a predefined set of distances $d(i,\omega)$. This procedure records first the unique magnitudes of the exhibited weights (i.e., radii $d(i,j)$) in the graph and for each such magnitude finds the minimum number of boxes required to cover all nodes in the graph. To minimize the computational (search) time for the minimum number of boxes, we use a greedy heuristic, which for each magnitude of the box prunes the original weighted graph by removing the edges that exceed the magnitude and clusters the nodes that are connected by weighted edges smaller than the current magnitude.  The algorithm then proceeds by analyzing and covering each cluster in descending order of their size (number of nodes). Knowing the number of boxes required to cover the weighted graph for each magnitude of the box allows us to investigate the multi-fractality and determine (estimate) the generating function of the counting measure as a function of the box radius.

Note that nodes that appear to be close to each other in the spin network domain representation may belong to different islands of concentration as a function of the adopted metric (e.g., information transfer distance).  For instance, while nodes $1$ and $2$ are neighboring to each other in the original spin network (see Fig.~\ref{f:MetricBasedDescription}), in the distance-based representation they may be further apart from each other. Although in Fig.~\ref{f:MetricBasedDescription} and throughout our current analysis we used only the information transfer metric defined above, the mathematical framework can be applied to other information-theoretic metrics and can be extended to analyze weighted graphs that can be generated by spin network interactions over time.

The multi-fractal analysis rests on a novel {\it node (counting) measure} defined as follows:
\begin{equation}
  \mu [ B_{k}(\epsilon) ] = \frac{N(\epsilon)}{N} = p_{k}(\epsilon), 
\end{equation} 
where $\epsilon$ represents the magnitude of the information distance, $B_{k}(\epsilon)= \{\omega \in \mathcal{G}: d(k,\omega) \leq \epsilon \} $ denotes a ball of radius $\epsilon$ centered in node $k \in \mathcal{V}$, $N$ is the total number of nodes in the spin network or weighted graph $\mathcal{G}$, and $N(\epsilon)$ denotes the number of nodes inside the ball $B_{k}(\epsilon)$ of radius $\epsilon$. Simply speaking, the node measure $\mu [ B_{k}(\epsilon) ]$ of a ball $B_{k}(\epsilon)$ (with radius $\epsilon$) centered in node $k$ represents the relative number of nodes that are covered by this ball or alternatively, the relative number of nodes whose ITF-distance to the node $k$ are smaller than $\epsilon$. Still in other words, $p_k(\epsilon)$ represents the probability  of finding the node $k$ in the ball $B_k(\epsilon)$.  Of note, this probability satisfies: $\sum_{k=1}^{N} [p_{k}(\epsilon)^{q}]_{q=1} = 1$ and $\sum_{k=1}^{N} [p_{k}(\epsilon)^{q}]_{q=0} = N$, respectively.  This counting measure bears similarities with and extends the multi-fractal formalism presented in~\cite{PhysRevLett.62.1327} such that the mass property is replaced by the counting of nodes covered by a ball of a certain radius on the graph.  A similar strategy can be generalized to hypergraphs, but this is left for future work.  

For a multi-fractal graph structure, the node counting measure satisfies the following relationship:
\begin{equation}
  \mu [B_{k}(\epsilon)] = c_{k,\alpha_{k}} \epsilon^{\alpha_{k}}, \mbox{ as } \epsilon \rightarrow 0,
\end{equation}
where $\alpha_{k}$ denotes the Lipschitz-H\"older exponent and $c_{k,\alpha_{k}}$ is a coefficient that depends on the box and the Lipschitz-H\"older exponent $\alpha_{k}$.  The Lipschitz-H\"older exponent can be defined for any measure $\mu$ and quantifies the singularity of the informational geometry.

The partition function can be expressed as:
\begin{equation}
  Z(q,\epsilon) = \sum\limits_{k=1}^{N} \{\mu [B_{k}(\epsilon)]\}^{q}
                      = \sum\limits_{k=1}^{N} \{ c_{k,\alpha_{k}} \epsilon^{\alpha_{k}} \}^{q},
\end{equation}
where $q$ is a real number with $q \in (-\infty,\infty)$ and the summation is upper bounded by $N$ representing the total number of boxes of size $\epsilon$.  By performing a histogram-like analysis (i.e., sorting and counting all terms corresponding to a particular Lipschitz-H\"older exponent $\alpha$), the partition function takes the following form:
\begin{equation}
 Z(q,\epsilon) = \sum\limits_{\alpha} \epsilon^{q\alpha} \sum\limits_{k \in \mathcal{G}_{\alpha}} c_{k,\alpha}^{q},
\end{equation}
where $\mathcal{G}_{\alpha}$ represents a subgraph characterized by a Lipschitz-H\"older exponent $\alpha$. The last summation term under the condition $q = 0$ represents the number of balls of radius $\epsilon$, i.e., $n_{\alpha} =\sum_{k \in \mathcal{G}_{\alpha}} c_{k,\alpha}^{0}$. For an infinitesimally small ITF $\epsilon$ (i.e., $\epsilon \rightarrow 0$) and a multi-fractal graph structure characterized by a Lipschitz-H\"older exponent $\alpha$, the number of balls of radius $\epsilon$ required to cover the entire embedding can be expressed as 
$$n_{\alpha} (=\mbox{number of balls or radius}\epsilon) = w(\alpha)\epsilon^{-f(\alpha)},$$ 
with $f(\alpha)$ denoting the multi-fractal spectrum. By corroborating these derivations, the partition function can be expressed as follows:
\begin{equation}
\label{e:f_of_alpha}
  Z(q,\epsilon) = \sum\limits_{\alpha} b(q,\alpha) w(\alpha)\epsilon^{q\alpha - f(\alpha)}, 
\end{equation}
where $b(q,\alpha) = n_{\alpha}^{-1} \sum_{k \in \mathcal{G}_{\alpha}} c_{k,\alpha}^{q}$ is a coefficient that depends on the number of balls of size $\epsilon$ required for covering the graph embedding.

\section{Mono-Fractal versus Multi-Fractal Distribution}

An alternative investigation of the multi-fractality of the information graph embedding can be done by studying the scaling behavior with respect to size $\epsilon$ and the $q$-dependence of the partition function $Z(q,\epsilon)$:
\begin{equation}
\label{e:alternative}
Z(q,\epsilon) = \sum\limits_{k=1}^{N} \{\mu [B_{k}(\epsilon)]\}^{q}
                      = g(q) \epsilon^{\tau(q)},
\end{equation} 
where $\tau(q)$ is called the mass exponent function and is used to quantify the scaling properties of the partition function. If the mass exponent $\tau(q)$ is a linear function of the $q$-dependent exponents then we call the distribution of node measure $\mu [B_{k}(\epsilon)]$ to be mono-fractal. Alternatively, if the mass exponent $\tau(q)$ is a nonlinear function of the $q$-dependent exponents then we call the distribution of node measure $\mu [B_{k}(\epsilon)]$ to be multi-fractal. 

Taking the multi-fractal spectrum $f(\alpha)$ in Eq.~\eqref{e:f_of_alpha} to be narrowly distributed,  at the limit $\epsilon^{-f(\alpha)}$ $\delta$-distributed, yields $\tau(q)$ in Eq.~\eqref{e:alternative} linear, hence mono-fractality.   

In addition, the mass exponent function $\tau(q)$ is related to the generalized dimension function $D(q)$ through the following equation: $\tau(q)=(q-1)D(q)$ or to the generalized Hurst exponent $H(q)$ through the following relation $\tau(q)=qH(q)-1$. Based on the above-mentioned arguments, a linear dependence of the mass exponent $\tau(q)$ implies that the generalized Hurst exponent $H(q)=H$ is independent of the $q$-dependent exponents. In contrast, a nonlinear dependence of the mass exponent $\tau(q)$ implies that the generalized Hurst exponent will also exhibit a nonlinearity with varying exponents $q$. 

\section{Comparison with Multi-fractal Anderson Localization}

Fluctuations around the metal-insulator criticality in Anderson localization is known to be  multifractal~\cite{anderson}.  While the fundamental mathematical techniques utilized in the latter are undoubtedly similar to ours,  here, it is applied to a situation significantly different from the Anderson localization.  First and most importantly, we do not deal with Anderson localization  (fast decay of eigenstate $\sum_k \Psi_k \ket{k}$);  we rather deal with maximum fidelity transfer between two states  ($|\langle j | \exp(-\imath Ht/\hbar)|i\rangle| \approx 1$),  without localizing the wave function at the terminal state $\ket{j}$. 

To make the comparison crisp, consider the partition function shared by both approaches:
\[ Z(q)=\sum_{k=1}^N \mu_k ^q = g(q)\epsilon^{\tau(q)}. \]
In~\cite{anderson}, the measure $\mu_k$ is $\sum_n |\Psi_n|^2$ for $n$ in the box $k$ of  physical size $\epsilon$,  while here $\mu_k$ is the relative number $N(\epsilon)/N$ of sites $n$ within the ``box" of  ITF size  $\epsilon$  centered at $k$.  To put it another way, setting $\mu_k=c_{k,\alpha_k}\epsilon^{\alpha_k}$, here $n_\alpha$ is the number of $\epsilon$ balls necessary to cover the network graph $\mathcal{G}$,  while in~\cite{anderson}, $n_\alpha$ is the number of sites such that  $\sum_{n \in\mathrm{Box} k}|\Psi_n|^2\sim\epsilon^\alpha$.    Clearly, the two counting measures are not the same,  nor can we use the counting measure of~\cite{anderson},  since eigenstates are irrelevant here.   We could think of using the transfer measure $1/d(i,j)$ to the trivial transfer from $i$ to $i$,  but this is bound to fail since $1/d(i,i)=1/0=\infty$.  Consequently, because of the irreconcilable difference between the counting measures,  the fractal spectra derived from $n_\alpha=\epsilon^{-f(\alpha)} $ will not be the same, not is their interpretation the same.  But probably the most significant discrepancy is the absence of disorder here, so crucial in Anderson localization. 

\section{Multi-Fractal Analysis of Chains}
\label{sec:chain}

One important spin network topology is represented by spin chains (see top left hand side of Figure \ref{f:MetricBasedDescription}).  To investigate the geometrical properties of a chain of spins, we use the information-metric-based mapping (see Fig.~\ref{f:MetricBasedDescription}) and estimate the partition function as a function of the order $q$ of higher order moments.  In this particular representation $q$ plays the role of the inverse of temperature from statistical thermodynamics. The observed statistical self-similarity (see Figure \ref{f:SpinNetworksSelfSimilarityIslands}) of the spin network translates into a power law relationship of the partition function and allows us to estimate the generalized Hurst exponent $H(q)$ and the multi-fractal spectrum $f(\alpha)$.

The rationale for investigating the generalized Hurst exponent $H(q)$ is motivated by the need to quantify the spatial heterogeneity that may exist in an information metric based representation. Simply speaking, we aim to study how small and large fluctuations across all node interactions contribute to particular patterns that may appear over multiple scales and influence the dependence of $H(q)$ as a function of order $q$. To some extent, we extend the meaning of generalized Hurst exponent from time series analysis and use it to quantify the roughness of the communication landscape. Consequently, in our framework, the generalized Hurst exponent represents a mathematical approach for investigating the scaling properties and measuring the degree of heterogeneity of the graph motifs over multiple spatial scales. More precisely, if the analysis of the $q$th-order moments of the distribution of information graph motifs shows no dependence on the generalized Hurst exponent with the order of the moment $q$, then the informational graph is considered homogeneous and mono-fractal. In contrast, if the generalized Hurst exponent exhibits significant variation over a wide range of $q$ orders, then the information graph is considered to be heterogeneous and multi-fractal. This multi-fractal structure (of the information based embedding of spin chains) can be understood as a divergence in terms of scaling trends between the short range (small fluctuations in ITF) and long-range (large fluctuations) ITF magnitudes.

\begin{figure*}[t]
\centerline{\includegraphics[width=7.2in,keepaspectratio]{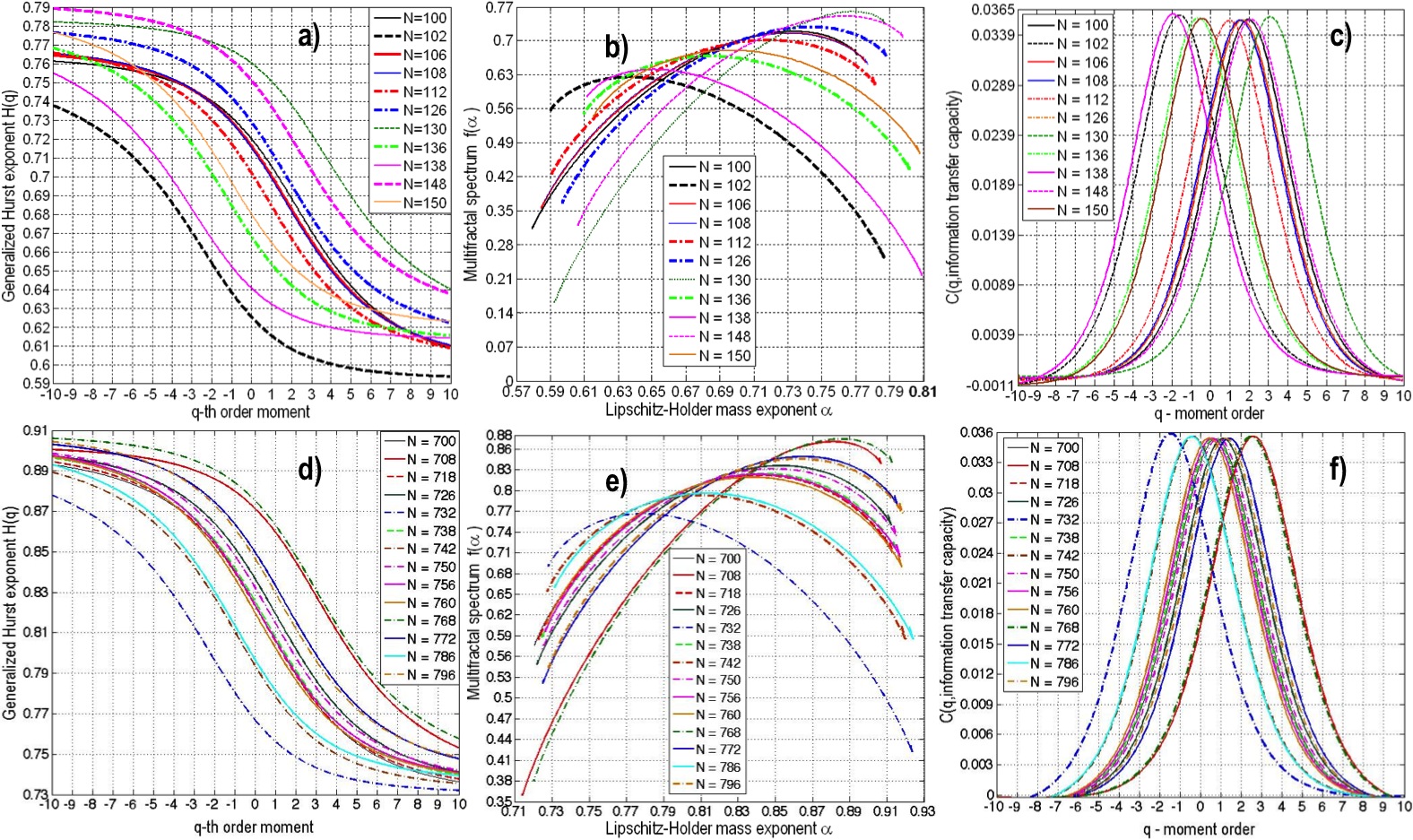}}
\caption{(a) Generalized Hurst exponent $H(q)$ as a function of $q$
  for several spin chain lengths (i.e., $N = 100, 102, 106, 108, 112,
  126, 130, 136, 138, 148, 150$) displaying a similar ``sigmoidal''
  shape, but rich multi-fractal behavior.
  (b) Multi-fractal spectrum $f(\alpha)$ as a function of the
  Lipschitz-H\"older mass exponent $\alpha$ for several spin chain
  lengths (i.e., $N = 100, 102, 106, 108, 112, 126, 130, 136, 138,
  148, 150$). Although spin chains exhibit similar multi-fractal
  spectrum, they are characterized by different dominant singularities
  (i.e., $\alpha$ at which $f(\alpha)$ attains maximum varies across
  spin chains).
  (c) Specific heat $C(q)$ for several lengths (i.e., $N = 100, 102,
  106, 108, 112, 126, 130, 136, 138, 148, 150$) of a spin chain.  (d)
  Generalized Hurst exponent $H(q)$ as a function of $q$ for several
  spin chain lengths (i.e., $N = 700, 708, 718, 726, 732, 738, 742,
  750, 756, 760, 768, 772, 786, 796$) displaying a similar
  ``sigmoidal" shape, but rich multi-fractal behavior.
  (e) Multi-fractal spectrum $f(\alpha)$ as a function of the
  Lipschitz-H\"older mass exponent $\alpha$ for several spin chain
  lengths (i.e., $N = 700, 708, 718, 726, 732, 738, 742, 750, 756,
  760, 768, 772, 786, 796$).
  (f) Specific heat $C(q)$ for several lengths (i.e., $N = 700, 708,
  718, 726, 732, 738, 742, 750, 756, 760, 768, 772, 786, 796$) of the
  spin chain. Although the length of the spin chain varies
  significantly, we observe similar multi-fractal patterns and
  curvature in the specific heat.}
\label{f:MultifractalAnalysisChain100and700}
\end{figure*}

Fig.~\ref{f:MultifractalAnalysisChain100and700}(a) shows the generalized Hurst exponent as a function of order $q$ for several spin chain lengths (i.e., $N = 100, 102, 106, 108, 112, 126, 130, 136, 138, 148$, and $150$).  The generalized Hurst exponent $H(q)$ displays a sigmoidal shape irrespective of the spin chain length.  Similar sigmoidal shapes are observed for numerous other spin chain lengths. Fig.~\ref{f:MultifractalAnalysisChain100and700}(d) summarizes the $H(q)$ vs $q$ dependency for spin chain lengths of $N=700, 708, 718, 726, 732, 738, 742, 750, 756, 760, 768, 772, 786$ and $796$.  This sigmoidal pattern shows that the information metric graph, having a heterogeneous architecture, is better characterized by multi-fractal geometric tools. Generally speaking, this implies that a single fractal dimension is insufficient to model the (heterogeneous) interactions and information transmission / propagation in the spin chain.

In addition to the sigmoidal shape, we also observe that for some lengths of the spin chain the generalized Hurst exponent exhibits a much more complex nonlinear dependency as a function of the $q$th order moment (see Figure \ref{f:MultifractalAnalysisChain105}(a) summarizing the analysis for spin chain lengths of $N = 105, 115, 119, 129$, and $149$). We also note that these particular spin chain lengths exhibit higher generalized Hurst exponents than those in Fig.~\ref{f:MultifractalAnalysisChain100and700}(a) and \ref{f:MultifractalAnalysisChain100and700}(d). This suggests that some spin chains exhibit a pronounced persistent behavior, i.e., a long interaction is likely to favor an even longer one, while others tend to display an anti-persistant behavior, interleaving short with long interactions.  From a practical perspective, it would be interesting to investigate the information processing / transmission properties of these two classes of spin chains on real devices, which may show some to be more suitable for information transmission while others might be better suited for robust information storage or parallel processing.

An alternative strategy for describing the local self-similar (scaling) properties of the information graph and quantify the degree of heterogeneity is to estimate the multi-fractal spectrum.  From a mathematical perspective, the multi-fractal spectrum represents the set of Lipschitz-H\"older exponents (fractal dimensions) and their likelihood of appearance as dictated by the mixture of locally self-similar motifs in the informational graph. Consequently, by estimating and analyzing the multi-fractal spectrum we can learn the existing dominance of some Lipschitz-H\"older exponents over others.  The Lipschitz-H\"older exponent quantifies the local singularities and locates the abrupt changes in the curvature of information graph embedding. More precisely, the maximum of the multi-fractal spectrum represents the dominant fractal dimension while the width of the spectrum is a measure of the heterogeneity richness (range of fractal dimensions) and complexity. From a structural point of view, the multi-fractality implies that the information graph consists of regions of short interactions (short information transmission ranges) mixed / interspersed with long-range interactions. Fig.~\ref{f:MultifractalAnalysisChain100and700}(b) and \ref{f:MultifractalAnalysisChain100and700}(e) show the multi-fractal spectrum for several spin chain lengths.  We observe that even though the generalized Hurst exponent for all these chain lengths exhibits a similar sigmoidal shape, the multi-fractal spectrum display different and asymmetric behavior. For instance, the multi-fractal spectrum of the spin chain of length $N = 102$ is prolonged over higher Lipschitz-H\"older exponents and thus stronger singularities, while the multi-fractal spectrum of the chain of size $N = 130$ is extending more towards lower Lipschitz-H\"older exponents corresponding to lower singularities in the curvature of the informational embedding.

The existence of these singularities in the information embedding suggests building on the multi-fractal analysis to develop a thermodynamic formalism of information propagation through the spin networks.  Of note, this thermodynamic formalism is not aimed at quantifying fluctuations over time but rather the spatially self-similar behavior in information transfer through a spin network. To elucidate the existence of a phase transition, we investigated the behavior of a thermodynamics inspired specific heat derived from the estimated partition function. Figures \ref{f:MultifractalAnalysisChain100and700}(c) and \ref{f:MultifractalAnalysisChain100and700}(f) show that the specific heat exhibits a bell shape whose peak values occur for various orders of $q$. In contrast, the specific heat for spin chains of lengths $N = 105, 115, 119, 129$, and $149$ in Figure \ref{f:MultifractalAnalysisChain105}(b) display a much more complex behavior that seems to be discontinuous in the vicinity of order $q=0$.

\section{Multi-Fractal Analysis of Rings}
\label{sec:rings}

\begin{figure*}[t]
\centerline{\includegraphics[width=7.2in,keepaspectratio]{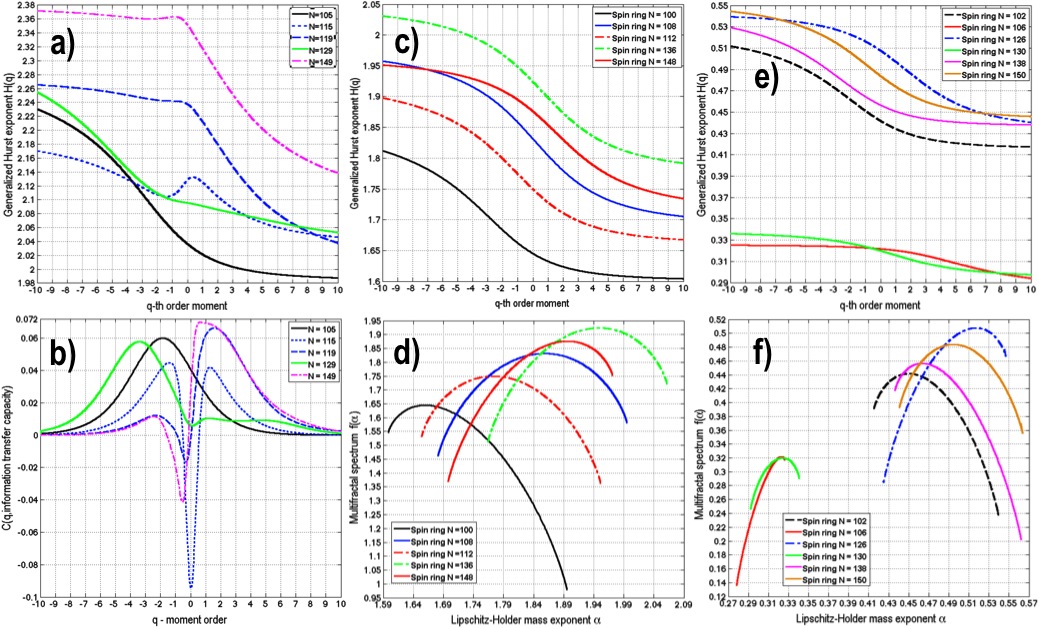}}
\caption{a) Generalized Hurst exponent $H(q)$ as a function of $q$ for
  spin chain lengths of $N = 105, 115, 119, 129$, and $149$ display a
  highly nonlinear behavior corresponding to rich multi-fractality. b)
  Specific heat $C(q)$ for spin chain lengths of $N = 105, 115, 119,
  129$, and $149$ exhibits a rich behavior that could be correlated to
  either a first- or a second-order (informational) phase
  transition. c) Generalized Hurst exponent $H(q)$ as a function of
  $q$ for several spin ring lengths (i.e., $N = 100, 108, 112, 136$,
  and $148$) displaying a similar ``sigmoidal" shape. e) The
  generalized Hurst exponent $H(q)$ as a function of $q$ for spin ring
  lengths of $N = 102, 126, 130, 138,$ and $140$ display a more
  pronounced multi-fractal behavior than for sizes of $N = 106$ and
  $130$. f) Comparison in terms of multi-fractal spectrum (shape and
  width) between spin ring networks of size $N = 102, 106, 126,
  130, 138, 140$ and $150$ .}
\label{f:MultifractalAnalysisChain105}
\end{figure*}

While spin chains have been most intensively studied in recent years, other arrangements such as rings also play an important role for quantum spintronic applications as their translation invariance properties make them potentially suitable as routers for quantum networks.  Consequently, it is important to study their information propagation characteristics. We have estimated the ITF metric for several ring sizes (i.e., from $N=50,\ldots,1000$), mapped the information propagation between all distinct pair of nodes $|i\rangle$ and $|j\rangle$, and applied our strategy for estimating the partition function over the metric graph. Fig.~\ref{f:MultifractalAnalysisChain105}(c) and \ref{f:MultifractalAnalysisChain105}(e) summarize the generalized Hurst exponents as a function of the $q$th order moment for several spin ring lengths. One observation we could make is that while the generalized Hurst exponents in Fig.~\ref{f:MultifractalAnalysisChain105}(c) are higher and reminiscent of a persistent dynamics, the generalized Hurst exponents in Figure \ref{f:MultifractalAnalysisChain105}(e) are much smaller (a third of those in Figure \ref{f:MultifractalAnalysisChain105}(c) and closer to 0.5) which would indicate a tendency of an anti-persistent structure. This distinction is even more interesting as it appears between spin rings of similar sizes. We have analyzed all spin rings $N = 50$ to $1000$ and we observed similar patterns. Consequently, it would be important to quantify the design implications and transmission properties against realistic devices.

Another observation we could make from Fig.~\ref{f:MultifractalAnalysisChain105}(c) and \ref{f:MultifractalAnalysisChain105}(e) is that most of the spin rings display a similar sigmoidal shape observed also for some spin chains, but a few spin rings (e.g., for $N=106$ and $130$) exhibit a generalized Hurst exponent that varies very little with order $q$. This can also be seen from the multi-fractal spectrum plots in Figures \ref{f:MultifractalAnalysisChain105}(d) and \ref{f:MultifractalAnalysisChain105}(f). For spin rings of sizes $N=106$ and $130$, we could conclude that are more closer to displaying a mono-fractal behavior (due to their lower generalized Hurst exponents and shrink multi-fractal spectrum).

\begin{figure*}[t]
\centerline{\includegraphics[width=7.2in,keepaspectratio]{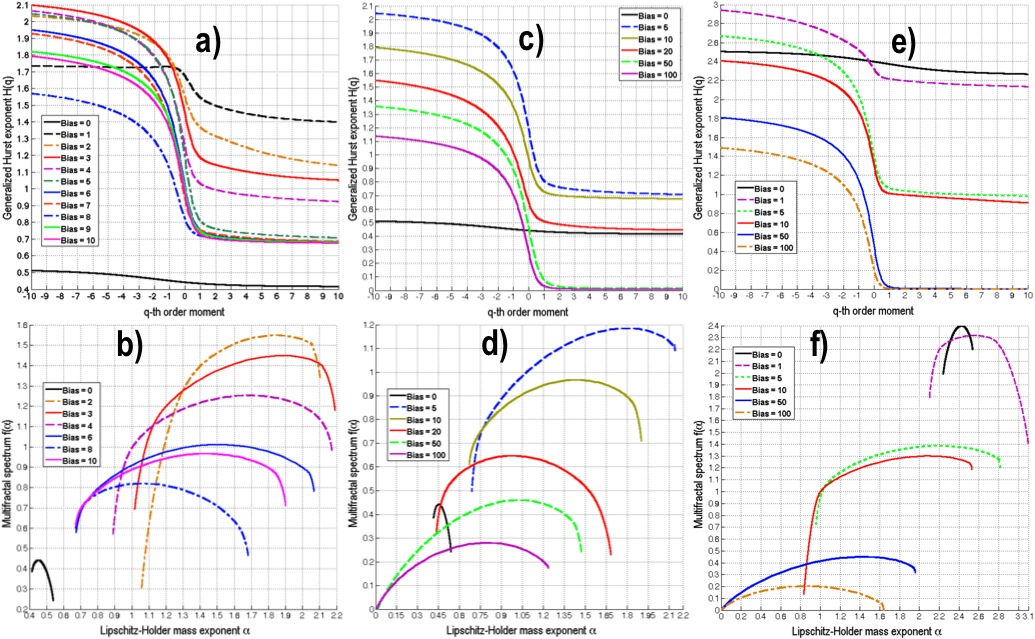}}
\caption{(a) Generalized Hurst exponent (GHE) $H(q)$ as a function of
  $q$ for spin ring of length $N=102$ and a bias $B$ of $0, 1, 2, 3,
  4, 5, 6, 7, 8, 9$, and $10$ applied to node $100$. The GHE displays
  a highly nonlinear behavior for non-zero bias B which corresponds to
  a richer multi-fractality.
  (b) The multi-fractal spectrum for a ring of length $N = 102$ and a
  bias $B$ of $0, 2, 3, 4, 6, 8$ and $10$.
  (c) The GHE $H(q)$ as a function of $q$ spin ring of length $N =
  102$ and a bias $B$ of $0, 5, 10, 20, 50$, and $100$.
  (d) The multi-fractal spectrum for the ring of size $N = 102$ and a
  bias $B$ of $0, 5, 10, 20, 50$ and $100$.
  (e) The GHE $H(q)$ as a function of $q$ for a spin ring of length $N
  = 500$ and a bias $B$ of $0, 1, 5, 10, 50$, and $100$.
  (f) The multi-fractal spectrum for a spin ring of length $N = 500$
  and a bias $B$ of $0, 1, 5, 10, 50$, and $100$.}
\label{f:BiasAnalysis}
\end{figure*}

Spin rings are of interest as they can act as quantum routers in internet-on-a-chip architectures.  Rings can be ``quenched" by applying a very strong bias (magnetic field) on one spin;  this has the effect of favoring transmissions symmetric relative to the bias and at the limit of infinte bias the ring is ``quenched" to a chain  with perfect information transfer fidelity between nodes next to the quench node.   Consequently, we studied the multi-fractal characteristics of spin rings as a function of the magnitude of applied bias. For instance, Figure \ref{f:BiasAnalysis}(a) and \ref{f:BiasAnalysis}(b) summarize the generalized Hurst exponent and the multi-fractal spectrum obtained for an information-metric embedding of a spin ring of size $N = 102$ when the bias is applied to node $100$ (the magnitude of the bias is assumed to take integer values between $0$ and $10$). We observe from Figure \ref{f:BiasAnalysis}(a) that with increasing bias the generalized Hurst exponent is shifting towards higher values. We also notice that the increase in the width of the generalized Hurst exponent is not monotonic with increasing bias magnitudes. This multi-fractal trend (of higher multi-fractality for non-zero bias) is also confirmed by the multi-fractal spectrum plot in Figure \ref{f:BiasAnalysis}(b). To further investigate this behavior, Figures \ref{f:BiasAnalysis}(c) and \ref{f:BiasAnalysis}(d) summarize the generalized Hurst exponent and the multi-fractal spectrum of the same configuration (spin ring of size $N = 102$) for bias magnitudes of $0, 5, 10, 20, 50$ and $100$. We can notice that with increasing bias magnitudes the support of the multi-fractal spectrum shifts towards left an anti-persistent region. To contrast these results, Figures \ref{f:BiasAnalysis}(e) and \ref{f:BiasAnalysis}(f) show the generalized Hurst exponent and the multi-fractal spectrum for a spin ring of size $N = 500$ and the bias is applied at node $100$. We observe that while the generalized Hurst exponent is significantly wider for nonzero bias, the multi-fractal spectrum in Figure \ref{f:BiasAnalysis}(f) displays a similar tendency of shifting towards left. We suspect that the multi-fractality of both original and ``quenched" spin rings is not only affected number-theoretic properties of the ring but also by the bias magnitudes. We plan to investigate these suppositions both analytically and experimentally in the future. 
    
\section{Discussion}
\label{sec:conclusions}

It is our belief that the complex ITF behavior exhibited by spin networks is a matter of their scaling properties.  By analyzing spin chains and rings up to size 1000 we observe that they display complex mono-fractal/multi-fractal structure depending on their symmetry.   In addition, a thermodynamics inspired framework reveals that several such spin chains and rings exhibit some forms of phase transition which could prove fundamental in the design of these future devices. 

The spin chain results suggest that information transmission in spin networks not only exhibits complex multi-fractal behavior but also nontrivial dependence on the size of the network.  Previous work on ITF bounds  and their attainability~\cite{Jonckheere2014,QIP2015} has shown that the size of the network, $N$, and various number theoretic issues play a role.  Our multi-fractal analysis appears to corroborate the observations made in prior studies and open the possibility to establish a connection between these fields.   In addition to $N$, it was shown in~\cite{Jonckheere2014} that the ITF depends on the greatest common divisor $\mathrm{gcd}(i,j)$,  where $i$ and $j$ are the input and output spins, respectively.  The somewhat repetitive pattern of $\mathrm{gcd}(i,j)$ may be the root cause underpinning  the multi-fractality.  

\begin{figure*}[t]
\centerline{\includegraphics[width=7.2in,keepaspectratio]{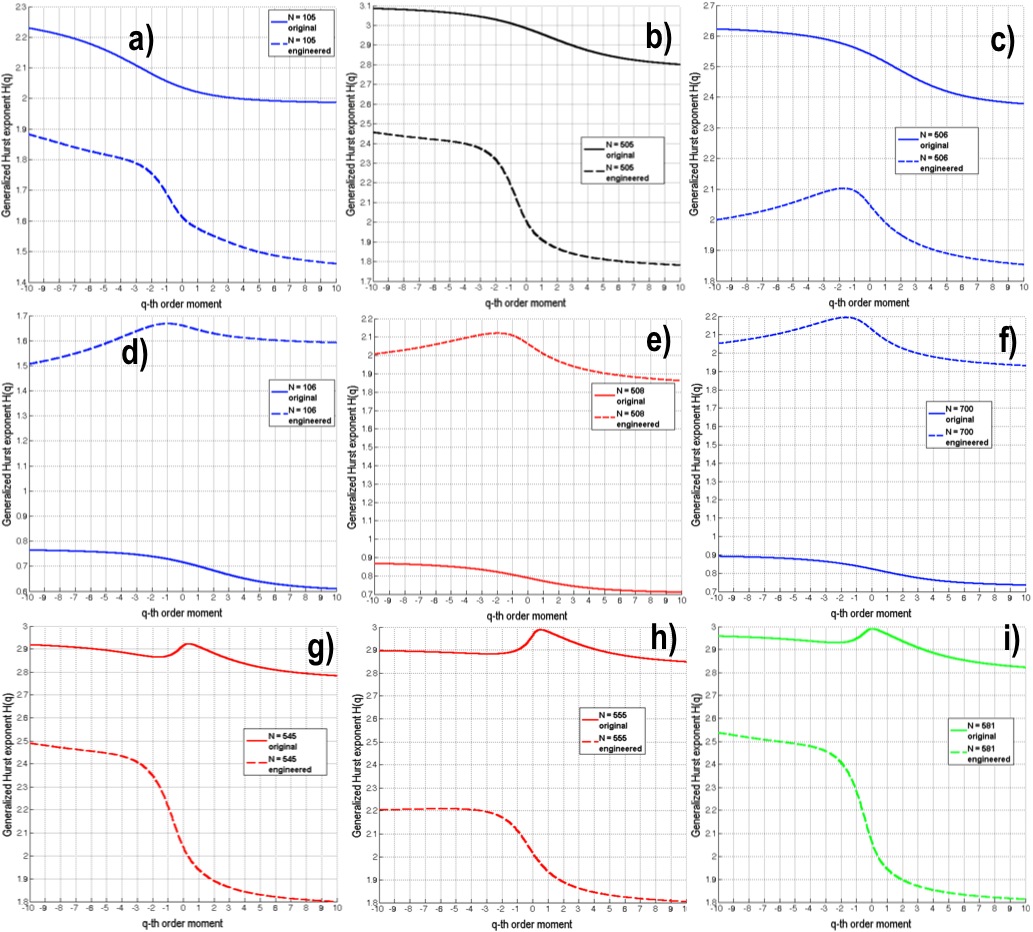}}
\caption{Comparison between original and engineered spin chains in
  terms of generalized Hurst exponent $H(q)$ for various chain
  lengths: a) $N = 105$, b) $N = 505$, c) $N = 506$, d) $N = 106$, e)
  $N = 508$, f) $N = 700$, g) $N = 545$, h) $N = 555$, and i) $N =
  581$.}
\label{f:ChainEngineeredResults}
\end{figure*}

In a near technological future,  chain will probably have their coupling strength engineered  to favor specific transfers.  It would be interesting to compare the multi-fractal properties such as the width and shape of the generalized Hurst exponent of engineered chains with those of a chain with uniform coupling.  Figure \ref{f:ChainEngineeredResults} shows some interesting trends from which we draw the following observations:
\begin{itemize}
\item The engineered chains have higher multi-fractality as evidenced by the increasing width of the generalized Hurst exponent.
\item The spin chains exhibit a dichotomous behavior in the sense that as a result of applying an engineered strategy in some cases it makes the generalized Hurst exponent shift above the original one while in other cases it shifts below the original one.
\end{itemize}

As far as rings are concerned, for zeros bias, they show a very narrow multi-fractal spectrum,  revealing that the symmetry contribute to mono-fractality.  However, rings can be endowed with  a non-trivial potential landscape  that changes the onsite potentials, corresponding to the diagonal elements in the Hamiltonian. As the bias increases we observe a higher degree of multi-fractality, consistent with the shape of the multi-fractal spectrum  of the engineered spin chains. 

The ITF developed here is an upper bound that becomes relevant  when enough time is given for the $i$ to $j$  transfer  to achieve its maximum fidelity. As already emphasized in~\cite{Jonckheere2014,QIP2015},  $|\langle j|\exp(-\imath H t/\hbar)| i\rangle|$ exhibits a complex, somewhat repetitive time dependence. This indicates that a spatio-temporal fractal analysis is warranted, but this is left for further research. 

Last but not least, the mathematical formalism constructed in this paper is general and can be applied to other metric-based networks---e.g., wireless networks.  Indeed, the very definition of the metric~\eqref{eq:d} is inspired from wireless networking~\cite{wireless},
\[ d(i,j)=-\log \mathrm{PRR}(i,j), \]
where $\mathrm{PRR}(i,j)$ is the Packet Reception Rate,  the probability of successful transmission of packets from node $i$ to node $j$.  Clearly, the mathematics underlying the two problem are indistinguishable, but the numerics would diverge as wireless networks lack long range dependence. In both spin and wireless networks curvature plays a crucial role~\cite{wireless},  with the difficulty of defining a curvature at all scales. Gromov developed a very large scale curvature, but of limited practicality for finite networks.  Probably the only way to resolve this difficulty is to define the curvature $\mathcal{G}_\alpha$ at every scale $\alpha$. 

\begin{acknowledgments}
P.B acknowledges support from National Science Foundation (NSF) under the 1453860 and 1331610 grants.  E. A. Jonckheere was partially supported by the Army Research Office (ARO) Multi University Research Initiative (MURI) grant W911NF-11-1-0268.  S. G. Schirmer acknowledges support from the Ser Cymru National Research Network on Advanced Engineering and funding from a Royal Society Leverhulme Trust Senior Fellowship.
\end{acknowledgments}

\end{document}